\documentstyle[aps,prl]{revtex}
\begin{document}
\input epsf.sty
\twocolumn[\hsize\textwidth\columnwidth\hsize\csname %
@twocolumnfalse\endcsname
\draft
\widetext


\title{Multiphase segregation and metal-insulator transition in
single crystal La$_{5/8-y}$Pr$_y$Ca$_{3/8}$MnO$_3$}

\author{V. Kiryukhin, B. G. Kim, V. Podzorov,
S-W. Cheong}
\address{Department of Physics and Astronomy, Rutgers University,
Piscataway, New Jersey 08854}
\author{T. Y. Koo}
\address{Bell Laboratories, Lucent Technologies, Murray Hill, New Jersey 07974}
\author{J. P. Hill}
\address{Department of Physics, Brookhaven National Laboratory, Upton,
New York 11973}
\author{I. Moon, Y. H. Jeong}
\address{Department of Physics, Pohang University of Science and Technology,
Pohang, Kyungbuk, 790-784, South Korea}

\date{\today}
\maketitle

\begin{abstract}

The insulator-metal transition in single crystal 
La$_{5/8-y}$Pr$_y$Ca$_{3/8}$MnO$_3$ with $y\approx$0.35 was studied using
synchrotron x-ray diffraction, electric
resistivity, magnetic 
susceptibility, and specific heat measurements.
Despite the dramatic drop in the resistivity at the insulator-metal
transition temperature T$_{MI}$, 
the charge-ordering (CO) peaks exhibit no anomaly
at this temperature 
and continue to {\it grow} below T$_{MI}$.
Our data suggest then, that in addition
to the CO phase, another insulating phase 
is present below T$_{CO}$. In this picture,
the insulator-metal transition is due to the changes
that occur {\it within} this latter phase. The CO phase does not appear to
play a major
role in this transition. We propose that a percolation-like
insulator-metal transition occurs
via the growth of ferromagnetic metallic domains within the parts of the
sample that do not exhibit charge ordering.  
Finally, we find that the low-temperature phase-separated 
state is unstable against x-ray irradiation,
which destroys the CO phase at low temperatures.

\end{abstract}


\phantom{.}
]
\narrowtext

\section{Introduction}

Metal-insulator transitions in manganite perovskites have attracted considerable
attention during the last five years \cite{Review}. It has been 
established that the metallic
state in these materials is ferromagnetic (with the double exchange mechanism
responsible for the ferromagnetism), and a variety of insulating states have
been found. In many cases, application of a 
magnetic field converts the insulating
phase into the ferromagnetic metallic (FM) state, resulting in the phenomenon of
``colossal magnetoresistance'' (CMR). Recently, it has been demonstrated
that microscopic phase separation plays an essential role in the physics of
the manganites \cite{Uehara,Casa,Kim,Babushkina}. 
In particular, it results in the apparent percolative character of the
insulator-metal transition when the transition is
from the charge-ordered insulating to the ferromagnetic metallic state. 
It is here that the largest changes in resistivity (more than 6 orders of
magnitude), and therefore the largest magnetoresistances are observed.

In spite of the large amount of work devoted to the manganites, the microscopic
nature of the phase-separated states has thus far not been understood. For
example, the electronic
properties of the constituent phases as well as their volume fractions
and spatial
distributions in the sample remain to be characterized. More importantly, the
physical mechanism underlying the phase separation phenomenon remains unclear.
It has been proposed theoretically that doped Mott insulators, such as 
the mixed-valence manganites, are unstable against electronic phase  
separation into carrier-rich and carrier-poor regions \cite{Yunoki}. 
This scenario,
however, is inconsistent with the sub-$\mu$m domain size observed experimentally
in several manganite materials \cite{Uehara}. 
More generally, it is unclear how such a
large-scale phase separation can be ascribed to the effects of short-range
local interactions or be consistent with long range Coulomb forces, 
and therefore other effects, including lattice strain
\cite{Littlewood} or quenched disorder \cite{New},
for instance, need to be considered. 
To address these basic questions and to understand the origin of the CMR
effect, the microscopic structure of manganites must be characterized in
detail, and, evidently, more experimental work is needed.
 
In this paper, we report synchrotron x-ray diffraction, electrical resistivity,
magnetization, and specific heat measurements 
performed on single crystal samples
of La$_{5/8-y}$Pr$_y$Ca$_{3/8}$MnO$_3$, $y$$\approx$0.35. 
With decreasing temperature,
this material first undergoes a charge-ordering (CO) transition at 
T$_{CO}\approx$200 K, and then a relatively 
sharp insulator-metal transition into a 
low-temperature conducting phase at T$_{MI}\approx$70 K. 
This low-temperature phase is believed to consist of a mixture of
charge-ordered insulating and ferromagnetic metallic 
phases \cite{Uehara}.
We estimated that
the low-temperature volume fraction of the ferromagnetic phase 
was 30\%
in this sample.
Note that the charge ordering is of the simple checker-board type found
in the ``$x$=0.5'' samples and therefore cannot be complete in the entire
sample at this doping.
We find that neither
the fraction of the CO phase in the sample, nor the correlation length of
the CO phase, show measurable
anomalies at T$_{MI}$. Moreover, the intensities of the CO diffraction peaks
continue to grow as the temperature is decreased across the
insulator-metal transition. Thus, 
these data show that the volume fraction of the CO phase does not decrease,
and in all likelihood continues to increase, as the material undergoes the
insulator-metal transition. Our combined data 
therefore indicate that in addition to 
the CO phase, another paramagnetic
insulating phase is present below T$_{CO}$. The 
insulator-metal transition is then caused by the changes {\it within} this
latter phase.
 
Existing experimental data suggest that 
the insulator-metal transition has a percolative character. We propose here that
it occurs
via the growth of ferromagnetic metallic domains within the parts of the
sample that do not exhibit charge ordering. 
In this picture there are at least three phases in our samples  
below T$_{CO}$: ferromagnetic metallic, 
charge-ordered insulating, and a second insulating state which is
paramagnetic above T$_{MI}$, and which we will refer to as the 
I2 phase (insulating phase 2). 
The mechanism of the metal-insulator transition in charge-ordered
manganites, therefore, appears to be more complex than a simple
percolation of metallic phase due to the growth of ferromagnetic
domains in the charge-ordered matrix.

In addition, we have studied the volume fraction of the FM phase 
in several other samples with $y\approx$0.35. The 
very low volume fractions of the FM phase observed in
some cases indicate that the FM phase is of a filamentary character 
even at the lowest temperatures. We argue therefore that 
crystallographic or magnetic domain boundaries, lattice
defects and associated strains may play an important role in the formation
of the conducting state in these materials.

Finally, we have investigated the
effects of x-ray irradiation on the low-temperature
phase-separated state.
As was the case in some other charge-ordered manganites 
\cite{Kir1,Kir2,Cox}, 
the CO state in our samples is destroyed by x-ray irradiation below T=50 K. 
The photoinduced transition was previously
found to be of the insulator-metal type
in related samples \cite{Kir1}. While the CO phase remains
unaffected by the photoinduced insulator-metal 
transition at temperatures larger than T$_{MI}$ in the related
(Pr,Ca,Sr)MnO$_3$ samples \cite{Casa}, the CO phase in the present case 
is destroyed by x-rays
at low temperatures. 
Thus, while an additional phase is required
to explain the insulator-metal transition at T$_{MI}$,
the low-temperature photoinduced
transition need only involve the FM and the CO phases. 

\section{Experiment}

Single crystals of La$_{5/8-y}$Pr$_y$Ca$_{3/8}$MnO$_3$ were grown using the
floating zone technique from polycrystalline rods with a nominal
composition near y=0.35 synthesized by a standard
solid state reaction method. 
Resistivity measurements were performed using a standard 
four-probe method, magnetization measurements were carried out with a
commercial SQUID magnetometer, and specific heat measurements were performed
using a recently developed Peltier microcalorimeter \cite{Peltier}. 

The x-ray diffraction measurements were carried out at beamline X22A
at the National Synchrotron Light Source. The 10.35 keV x-ray beam was 
focused by a mirror, monochromatized by a Si (111) monochromator, scattered
from the sample mounted inside a closed-cycle cryostat, and analyzed with
a Ge (111) crystal. The x-ray beam was $\sim$0.5$\times$1 mm in
cross section, and the x-ray flux was $\sim$10$^{11}$ photons per second.
The typical mosaic spread in our samples
was 0.2$^\circ$. 

Below the charge-ordering transition temperature
T$_{CO}$, superlattice diffraction peaks of two types appear. The 
(H, K/2, L) peaks, K odd (in the orthorhombic {\it Pbnm} notation), 
are associated with the Jahn-Teller distortions 
characteristic of the CE-type charge and orbitally ordered
state, and are often referred to as orbital-ordering
peaks. The (H, 0, 0) and (0, K, 0) peaks, 
H and K odd, have the 
same wavevector as the checkerboard
charge ordering and arise from lattice distortions associated with valence
ordering.
In a recent series of experiments \cite{Zimm}, these reflections were studied 
in the related material Pr$_{1-x}$Ca$_x$MnO$_3$ which also exhibits the 
CE-type CO state. By tuning the incident x-ray energy through the Mn absorption
edge, characteristic resonance and polarization dependences were observed at
these reflections, directly confirming their assignment as orbital and
charge-ordering peaks. 
In our measurements
the x-ray energy is far from resonance, and therefore both types of superlattice
reflections arise from lattice distortions associated with
the CO state. 
In this work, we concentrated on the (0, 4.5, 0), (0, 5, 0), and
(0, 5.5, 0) peaks. Longitudinal (parallel to the scattering vector) and 
transverse scans were taken. The longitudinal scans were fitted
using Lorentzian-squared line shapes, and these fits were used to extract the
intensities, positions, and widths of the peaks.   

\section{Results and Discussion}

\begin{figure}
\centerline{\epsfxsize=2.9in\epsfbox{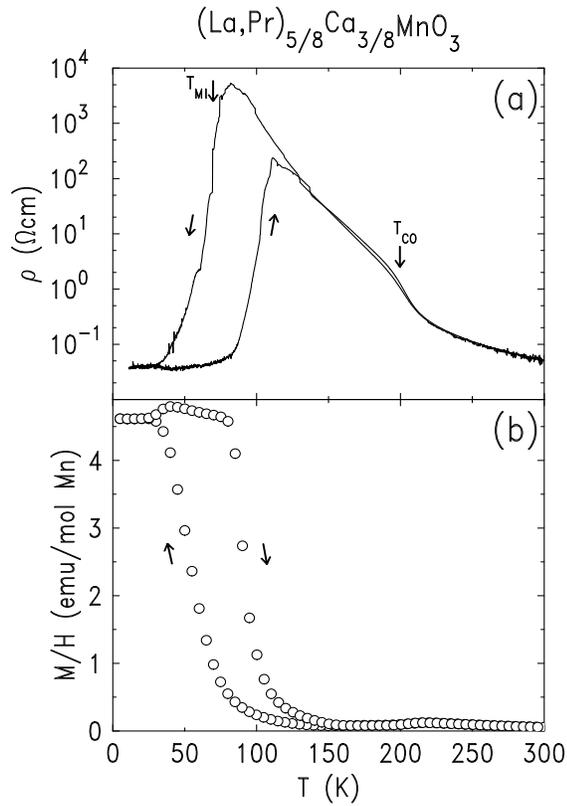}}
\vskip 5mm
\caption{Temperature dependences of (a) zero-field electric resistivity and
(b) the magnetic susceptibility in a 100 Oe magnetic field, taken on heating
and on cooling. The transition temperatures T$_{CO}$ and T$_{MI}$ were
determined from the maxima in the temperature derivative
of the logarithmic resistivity on cooling.} 
\label{fig1}
\end{figure}

Fig. \ref{fig1} shows the temperature dependences of the zero-field electric
resistivity 
and magnetization in a 100 Oe magnetic field. 
The anomalies at T$_{CO}\approx$200
K are due to the CO transition. With decreasing temperature, a 
relatively sharp insulator-metal transition 
occurs at T$_{MI}\approx$70 K. 
(The transition temperatures were defined as the temperatures of the
maxima in the temperature derivative of the logarithmic resistivity.)
In the vicinity of T$_{MI}$, the sample magnetization
gradually increases on cooling before saturating below T=40 K. The transition is
strongly hysteretic. 

\begin{figure}
\centerline{\epsfxsize=2.9in\epsfbox{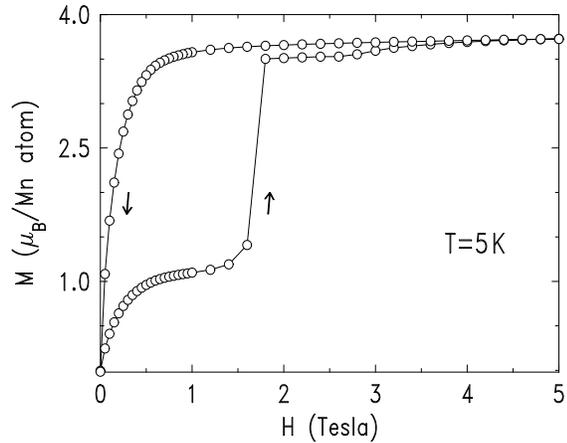}}
\vskip 5mm
\caption{Magnetization versus magnetic field in ZFC sample taken on ramping
the field up and down at T=5 K.}
\label{fig2}
\end{figure}

In a recent work by Uehara {\it et al.} \cite{Uehara}
it was shown that 
the low-temperature state of this material is inhomogeneous,
and that even at very low temperatures only a fraction
of the sample becomes metallic. Assuming that these metallic parts
of the sample are also ferromagnetic, this fraction can be estimated using
the data of Fig. \ref{fig2}, which shows the sample magnetization versus
applied magnetic field in a zero-field cooled sample. The magnetization first
saturates at a field of about 1 Tesla. This saturation is attributed to 
the complete alignment of the FM domains present in the sample 
in zero field \cite{Uehara}.
The material then undergoes two field-induced transitions at the fields of
1.6 and 3 Tesla. Finally,
when the field is increased to H=4 Tesla the 
entire sample becomes ferromagnetic, as indicated by the magnitude of the
saturated magnetic moment. The field-induced transition is persistent, and the
entire sample remains in the FM state after the field is turned off. Therefore,
the fraction of the FM state in the zero-field cooled sample can be determined
from the ratio of the low-field (H$<$1 Tesla)
magnetization taken on ramping the field up and
down. Using the data of Figs. \ref{fig1} and \ref{fig2}, 
we thus conclude that the
fraction of the FM phase in our sample was $\sim$30\% at T=5 K and
$\sim$9\% at T=70 K, the insulator-metal transition temperature.

One possible scenario for the insulator-metal transition in 
La$_{5/8-y}$Pr$_y$Ca$_{3/8}$MnO$_3$ involves the growth of the metallic
domains within the charge-ordered insulating matrix with decreasing
temperature. The system would then undergo 
the insulator-metal transition when these metallic domains percolated.
As discussed above, the volume fraction of the conducting phase
can be estimated from the magnetization measurements, and
in our samples should thus increase from approximately zero to $\sim$30\% as the
temperature decreases from 100 K to 40 K. 

\begin{figure}
\centerline{\epsfxsize=2.9in\epsfbox{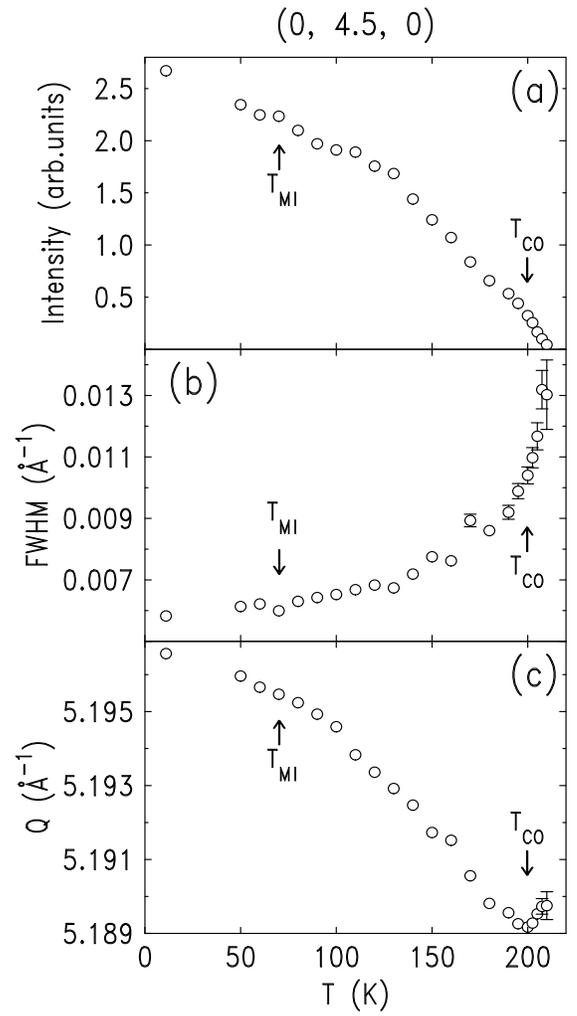}}
\vskip 5mm
\caption{Temperature dependences of (a) intensity, (b) full width at half
maximum, and (c) scattering vector magnitude of the (0, 4.5, 0)
superlattice diffraction peak. The data were taken on cooling.
The transition temperatures T$_{CO}$ and T$_{MI}$ were determined from the
resistivity data of Fig. 1.}
\label{fig3}
\end{figure}

To test this hypothesis, we investigated
the properties of the CO state using synchrotron x-ray diffraction.  
The temperature dependences of the intensity, width, and the scattering 
vector of the (0, 4.5, 0) superlattice peak are shown in Fig. \ref{fig3}.
The data were taken on cooling.
The intensity of this peak is often taken as the order parameter of the 
CE-type charge and orbital ordered state
because it reflects the degree of the order achieved in the 
CO system provided that the sample is homogeneous. In an inhomogeneous 
system, this intensity can also increase if 
the volume fraction of the CO phase increases. 
Therefore, if the conducting phase grows at the expense of the CO phase as the
temperature is decreased, one would expect the CO peak intensity to decrease 
as the sample undergoes the insulator-metal transition. 
In principle, it is possible that
the increase in the peak intensity due to
the improved ordering at low temperatures might compensate for any decrease
in the volume fraction of the CO phase.
However, taking into account the high
volume fraction (30\%) of the ferromagnetic
phase at low temperatures and noting that
the order parameter of the CO phase is not expected to change rapidly 
at temperatures much smaller than
T$_{CO}$, we believe that the latter possibility is very
unlikely \cite{saturation}. 
That is, if 30\% of the CO phase were to be transformed, we would be
able to observe it.
It is surprising therefore that the data of 
Fig. \ref{fig3}(a) do not show any decrease in the (0, 4.5, 0) peak intensity
in the vicinity of 
the insulator-metal transition. On the contrary, the peak intensity
continues to increase with decreasing temperature down to T=10 K,
showing no detectable anomaly at T$_{MI}$. The intensity of the (0, 5, 0)
CO peak also does not show any decrease in the vicinity of T$_{MI}$. 
We therefore conclude that
our data are
inconsistent with any model in which the insulator-metal transition is due to
the simple growth of the FM phase at the expense of 
the CO phase at low temperatures.

Panel (b) of Fig. \ref{fig3} shows 
the temperature dependence of the (0, 4.5, 0) peak width. The
intrinsic peak width is inversely proportional to the correlation length
of the CE-type ordered state. 
The data of Fig. \ref{fig3}b show that the correlation
length of the orbital
state remains finite well below the CO transition temperature,
slowly growing with decreasing temperature. Even at 10 degrees below T$_{CO}$,
it does not exceed 200 $\rm\AA$. It is unclear to what extent these data
reflect the size of the CO domains, however, since similar peak broadening
of orbital reflections
was observed in Pr$_{0.5}$Ca$_{0.5}$MnO$_3$ samples 
in which CO reflections exhibit long range order \cite{Zimm}.
In any case, it is evident that 
the orbital state is highly disordered in the vicinity of T$_{CO}$ but 
becomes progressively better correlated as the temperature decreases. The
CO domain size may also grow in the process. 
However, as was the case for the peak intensity,
there is no anomaly in the temperature dependence of the peak width in the
vicinity of T$_{MI}$, and, clearly, there is no low-temperature width
increase that might be expected if charge ordering is destroyed in a
third of the sample volume. 

The temperature dependence of the (0, 4.5, 0) peak position is shown in Fig. 
\ref{fig3}(c).
There is an abrupt change in
this temperature dependence at T=200 K. This change is almost certainly
caused by the structural transition. The increase in the CO scattering
vector above T=200 K may be caused by the corresponding decrease in the
$b$-axis
lattice constant, if the CO fluctuations simply follow the underlying crystal
lattice. However, more interesting explanations, including the possibility
of incommensurate orbital
fluctuations that were previously observed in other
manganites \cite{Incomm}, are also possible.

\begin{figure}
\centerline{\epsfxsize=2.9in\epsfbox{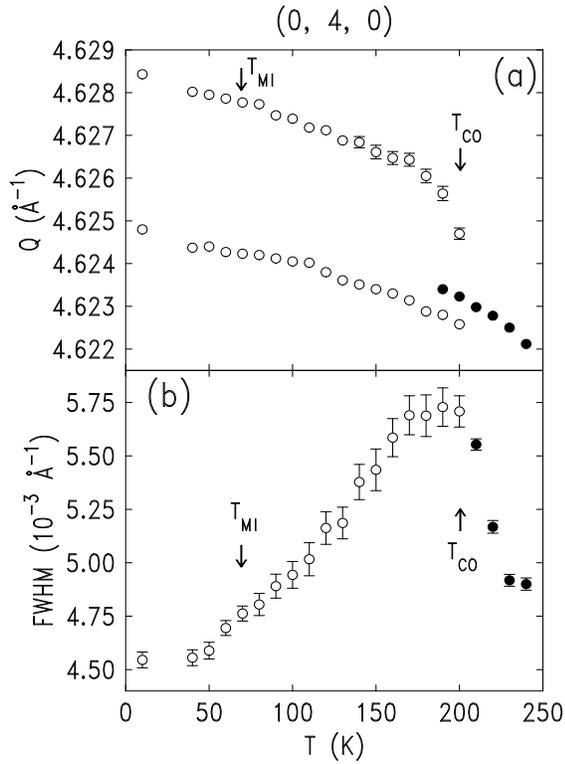}}
\vskip 5mm
\caption{Temperature dependences of (a) peak positions and (b) full width at 
half maximum of the Bragg peaks observed in the vicinity of the (0, 4, 0)
position in reciprocal space. The data were taken on cooling.
The data were fitted to either a single peak (filled circles) or two
overlapping peaks (open circles). The transition temperatures T$_{CO}$ and
T$_{MI}$ are from the data of Fig. 1.}
\label{fig4}
\end{figure}

We have also studied the
temperature-dependent behavior of the fundamental Bragg peaks. In particular,
scattering in the vicinity of the (0, 4, 0) position in the reciprocal space
was measured, and the results are shown in Fig. \ref{fig4}. Note that due to
the presence of twin domains, (0, 4, 0), (4, 0, 0), and (2, 2, 4) reflections 
may, in principle, be present at this reciprocal space position.
Similar to many
other charge-ordered manganites, our sample undergoes a structural transition
at T$_{CO}$. Above T=220 K a single narrow peak is observed, while
below T=180 K two overlapping peaks are present (see Fig. \ref{fig5} for
an example of the low-temperature peak profile). It was impossible to 
distinguish between a single broad peak or closely separated two peaks in the
vicinity of T$_{CO}$. Below T$_{CO}$, peak positions and widths evolve
in a smooth manner down to T=10 K. 
There appears to be
a small but systematic narrowing of the Bragg peaks with decreasing temperature
below T$_{CO}$. This decrease in the peak width indicates that lattice
strain is relieved as the temperature is decreased. The lattice strain is
maximized in the vicinity of T$_{CO}$. Note that some of the decrease of
the (0, 4.5, 0) peak width at low temperatures (Fig. \ref{fig3}(b)) may
thus be attributed to the decrease of the overall lattice strain. 

\begin{figure}
\centerline{\epsfxsize=2.9in\epsfbox{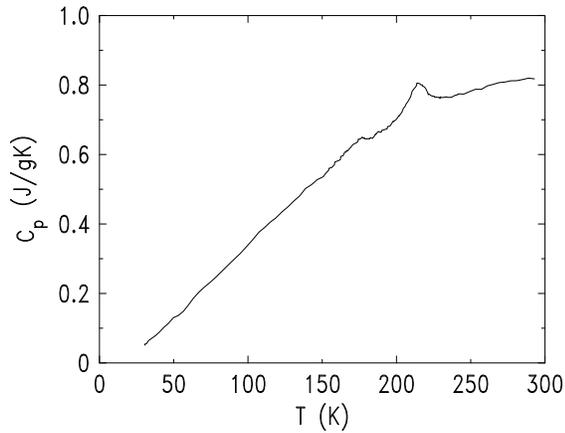}}
\vskip 5mm
\caption{Temperature dependence of the specific heat.}  
\label{heat}
\end{figure}

The temperature dependence of the specific heat is shown in Fig. \ref{heat}.
The temperature anomaly at T=210 K is due to the charge-ordering transition
\cite{difference}.
The anomaly at T=170 K is likely the result of the N\'eel transition which
takes place in Pr$_{1-x}$Ca$_x$MnO$_3$ samples at a similar temperature
\cite{Tomioka}. As was the case for
the diffraction data, the specific heat exhibits
no anomaly at T$_{MI}$. This behavior is consistent with the multiphase
scenario for the insulator-metal transition in our sample and is
clearly different from the behavior exhibited by
La$_{0.7}$Ca$_{0.3}$MnO$_3$. The latter compound has approximately the same
doping level and exhibits a conventional metal-insulator transition at
the Curie temperature at which a pronounced specific heat anomaly is observed 
\cite{Cp}. 
 
The combined data discussed above are clearly incompatible with the picture
in which the FM phase appears at a temperature of approximately 100 K, as
the magnetization data would suggest, and then grows at the expense of the 
CO phase as the temperature decreases, 
percolating at T$_{MI}$=70 K and finally reaching a volume fraction of 30\%
at T=40 K. On the contrary, the temperature dependences of Fig. \ref{fig3}
indicate that the correlation length, and possibly even the volume 
fraction of the CO phase,
grow with decreasing temperature. 

Based on our data, we therefore propose the 
following modified scenario for the insulator-metal transition in
La$_{5/8-y}$Pr$_y$Ca$_{3/8}$MnO$_3$. 
A secondary phase (the insulating phase 2, or the I2 phase), 
which is distinct from the phase that
undergoes the CO transition, is present in our samples below T$_{CO}$.
Since the CO phase is insulating and its volume fraction is not decreasing
with decreasing temperature,
the changes {\it within} this secondary phase must account for the 
insulator-metal transition.   
That is, the insulator-metal transition results from
the growth of ferromagnetic metallic domains within the parts of the
sample that do {\it not} exhibit charge ordering.

There are at least two possibilities for this growth. 
The FM domains can percolate
within the I2 phase in a manner previously suggested for the percolation
of metallic domains within the CO matrix.
Alternatively, the I2 phase can become a
ferromagnetic metal in a more or less uniform manner, possibly
with some small volume
fraction undergoing the transition at lower temperatures due to
local variations of lattice strain. The presence of this latter fraction
may lead to the formation of the insulating ``bottlenecks'' in an otherwise
conducting secondary phase near T$_{MI}$. Small fluctuations in these
insulating regions would then lead to large changes in the sample resistance.
In the both scenarios, the insulator-metal
transition would exhibit properties characteristic of the percolative
transition, as have indeed been found in a variety
of manganite samples \cite{Uehara,Casa,Kim,Babushkina}, with arguably the most
dramatic manifestation being the colossal fluctuations observed in $1/f$
noise measurements \cite{Podzorov,Molnar}. 
In summary, our data strongly suggest that the
insulator-metal transition in our sample is due to the changes that occur
within the non-charge-ordered parts of the sample.
We would like to note that very recently the existence of the secondary
low-temperature insulating phase was reported in Pr$_{0.7}$Ca$_{0.3}$MnO$_3$
samples \cite{Radaelli}.
A detailed crystallographic
study and structural refinement are needed to establish whether the I2 phase
in our sample and the secondary insulating phase in Pr$_{0.7}$Ca$_{0.3}$MnO$_3$
are the same.

It has
been recently found that in many cases the CO state in manganite
materials is unstable against irradiation with x-rays \cite{Kir1,Kir2} or even
with visible light \cite{Miyano}. In particular, in
Pr$_{0.7}$Ca$_{0.3}$MnO$_3$ x-rays destroy the CO state and convert the
material to a conducting state which was conjectured to be ferromagnetic
\cite{Kir1}.
More recently it has been
shown that the photoinduced state in related thin films does indeed possess
a substantial magnetic moment \cite{Baran}. 
We have also therefore carried out a study of the effects of x-ray irradiation
in our samples, in particular with the intention of
comparing the x-ray-induced transition with the temperature-induced
insulator-metal transition
that occurs at T$_{MI}$ in the absence of x-rays.

\begin{figure}
\centerline{\epsfxsize=2.9in\epsfbox{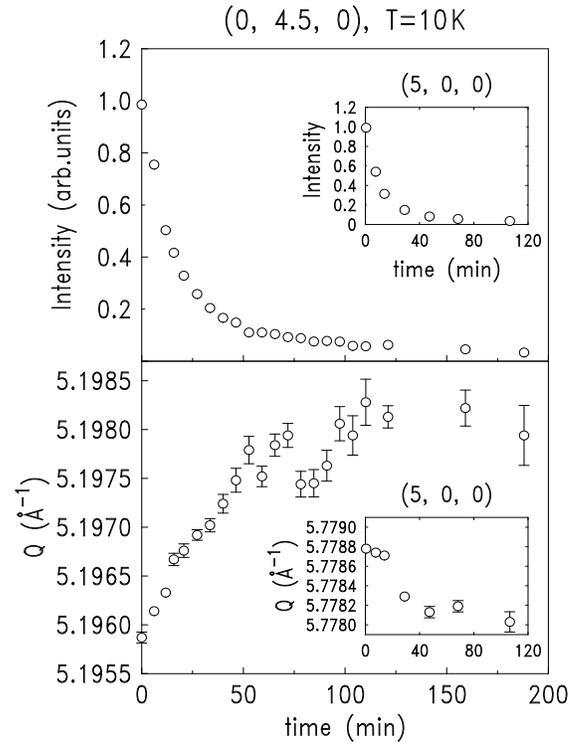}}
\vskip 5mm
\caption{X-ray exposure dependences of the (0, 4.5, 0) superlattice peak
intensity (top panel) and its position (bottom panel) at T=10 K. The insets 
show the corresponding dependences for the (5, 0, 0) charge-ordering peak.}
\label{fig6}
\end{figure}

We find that in our samples 
the charge ordering is also
destroyed by x-ray illumination at low temperatures. Fig. \ref{fig6} shows the
intensity and the scattering vector of the (0, 4.5, 0) superlattice peak
versus x-ray exposure time. Note that as the crystal lattice
gradually relaxes in the transition process, the position of the (0, 4.5, 0)
peak also changes, following the changes in the $b$ lattice
constant. A diffraction peak was also found at the
(5, 0, 0) position. This peak is present only in the CO state and
disappears as the sample is irradiated (insets in Fig. \ref{fig6}).
After prolonged x-ray irradiation, the $a$ lattice constant of the
remaining CO domains increases
by 0.01\% and the $b$ constant decreases by 0.045\%.

\begin{figure}
\centerline{\epsfxsize=2.9in\epsfbox{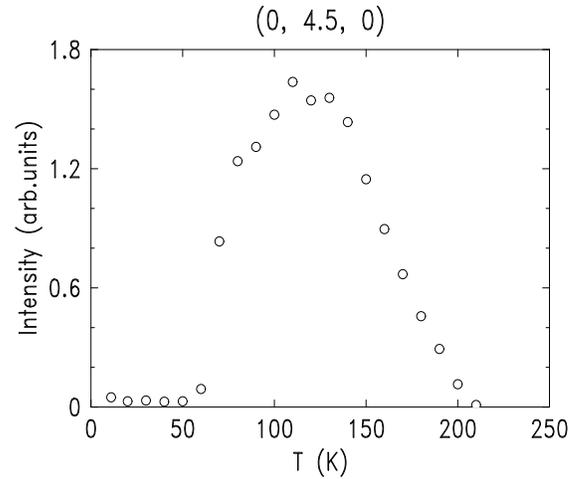}}
\vskip 5mm
\caption{The intensity of the (0, 4.5, 0) superlattice peak taken on 
heating from the state in which charge ordering was destroyed by long x-ray
exposure.} 
\label{fig7}
\end{figure}

\begin{figure}
\centerline{\epsfxsize=2.9in\epsfbox{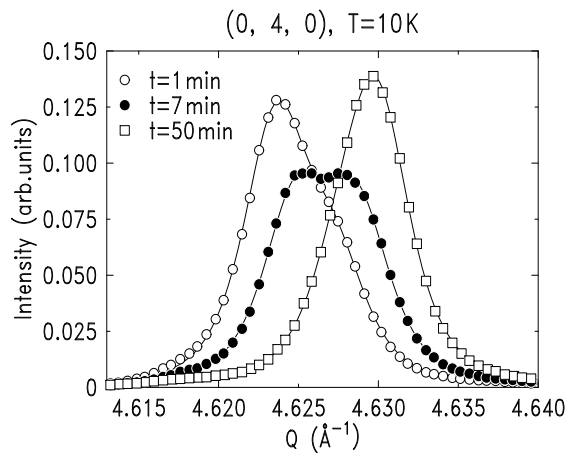}}
\vskip 5mm
\caption{Longitudinal diffraction scans taken in the vicinity of the (0, 4, 0)
position in reciprocal space after various x-ray exposures. The temperature
was T=10 K.}
\label{fig5}
\end{figure}

The x-ray-induced effects are present only at
temperatures less than 50 K. In fact, the CO state is recovered on
heating the x-ray converted samples above T=60 K, as shown in Fig. \ref{fig7}.
The x-ray induced FM state, therefore,
is unstable above T=60 K.
Similar phenomenology was observed previously in
Pr$_{0.7}$Ca$_{0.3}$MnO$_3$ samples \cite{Kir1}.
We note that the data of Fig. \ref{fig3} are unaffected by these
x-ray effects since we find no such effects at T=50 K and above
(on cooling), and the data of Fig. \ref{fig3}
at T=10 K were taken quickly after the sample was cooled from 50 K
to 10 K in the absence of x-rays.

Because of the strong electron-lattice coupling in the manganites \cite{Andy},
lattice parameters of the FM state are expected to be different from those 
of the CO phase. It is therefore not surprising that our samples undergo
substantial structural changes when irradiated with x-rays below T$\sim$50 K.
Fig. \ref{fig5} shows longitudinal scans (parallel to the scattering vector)
in the vicinity of the (0, 4, 0) allowed Bragg 
peak at T=10 K after various x-ray
exposures. The scans were collected using an attenuated beam,
so that the x-ray-induced change during the course of a single
measurement was negligible. 
The data of Fig. \ref{fig5} show that lattice constants of the x-ray-induced
phase differ substantially from those of the non-irradiated material. 
We have tried fitting these data assuming the presence of several phases with
fixed lattice constants, but the best fits were obtained when the lattice
parameters of the phases were allowed to vary. Such behavior could be the result
of a gradual relaxation of the lattice strain exerted by the CO phase
on the x-ray-induced phase as the latter phase grows within the CO matrix.

There appears to be a difference between the 
thermally-induced insulator-metal transition
at T$_{MI}$ in the absence of x-rays and the x-ray-induced
transition at low temperatures. In the former case, the CO phase is not
affected in any measurable manner, while in the latter the conductivity
increases simultaneously with the destruction of the CO phase 
in the related samples \cite{Kir1}, 
and therefore
the CO phase is almost certainly converted into the metallic state in the 
process. Other evidence that the two transitions have different mechanisms
comes from the observation that when the samples of a composition
similar to ours
are irradiated with x-rays {\it above} T$_{MI}$, the changes in the intensity
of the CO peaks are minimal or even absent \cite{Casa} while an
insulator-metal transition still takes place. Thus, it appears that
while a phase distinct from both the FM and the CO phases is required to
explain the transition at the relatively high transition temperature
T$_{MI}$, the {\it low-temperature} x-ray-induced
transition may involve only two phases. It would be very interesting to
check if this statement holds for other external perturbations, such as
magnetic field, for example. 

\begin{figure}
\centerline{\epsfxsize=2.9in\epsfbox{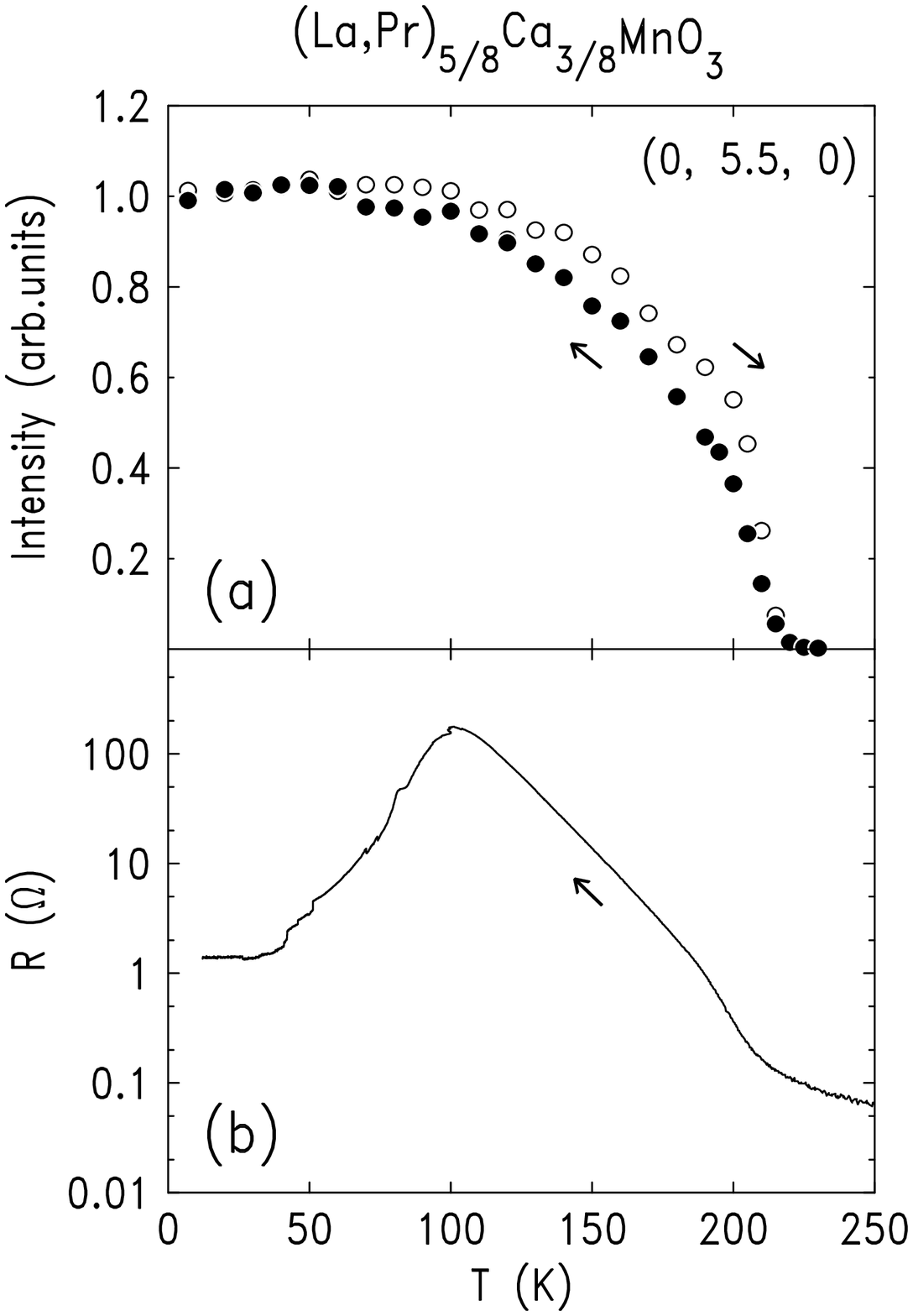}}
\vskip 5mm
\caption{Temperature dependences of (a) the intensity of the (0, 5.5, 0)
superlattice peak, and (b) the electric resistivity in the 
sample with a small low-temperature fraction of the FM phase.} 
\label{fig8}
\end{figure}

Finally, we briefly discuss samples of different compositions. 
The exact sample composition depends on the nominal
composition of the sample feed rod and on the preparation conditions.
Extensive studies on how the properties of ceramic samples of
La$_{5/8-y}$Pr$_y$Ca$_{3/8}$MnO$_3$ change with composition can be
found in Refs. \cite{Uehara,Kim}. 
We have investigated several samples with nominal $y\approx$0.35,
all of which exhibited the
insulator-metal transition at low temperatures. The magnitude of the
resistivity drop at this transition, as well as the low-temperature fraction 
of the FM phase determined as described above, were different in different
samples. Fig. \ref{fig8} shows temperature dependences of the 
(0, 5.5, 0) peak intensity
and the electric resistance in one of these samples. The 
low-temperature volume fraction
of the FM phase determined from magnetization measurements was only
2.5\%. Nevertheless, the sample resistance dropped by more than 2 orders
of magnitude on cooling from 100 K to 50 K. The temperature-dependent
behavior of the CO phase is perfectly conventional, 
that is, the intensities and correlation lengths of the charge and orbital
ordering saturate at low temperatures and no anomalies are found in the
vicinity of T$_{MI}$.
Therefore, the conducting phase in this particular
sample must be of a highly filamentary character. Otherwise,
a conducting path could not form because of the small fraction of the 
conducting phase in this sample. 

It is natural to assume that
local strains associated with twin-domain boundaries and other kinds of
structural defects can significantly change electronic properties of the
nearby regions
due to the strong electron-lattice coupling \cite{Andy}. 
The filamentary FM regions that would be induced by such extended lattice 
defects might appear to be a plausible explanation for the reduced
low-temperature resistivity in our samples which have a small low-temperature 
fraction of the FM phase. The important conclusion that can
be drawn from this is that electronic properties of manganites samples can,
in some cases, be strongly affected by extremely small amounts of a
secondary phase that can arise naturally due to structural defects. Great 
care in choosing sample growth conditions and sample preparation methods
should, therefore, be taken in order to study the properties of the
``pure'' phases.

\section{Conclusions}

In conclusion, we find that the simple percolation model of the 
insulator-metal transition in La$_{5/8-y}$Pr$_y$Ca$_{3/8}$MnO$_3$
needs to be modified. Our data are inconsistent with the growth of
the FM state at the expense of the CO state at low temperatures,
at least not in the amounts suggested by the magnetization measurements.
We propose that parts of the sample, while insulating,
do not exhibit charge ordering below
T$_{CO}$ and that the insulator-metal transition is due to the growth
of ferromagnetic metallic domains within these parts. 
Whether these metallic domains actually percolate at T$_{MI}$, or more 
complex structures containing insulating ``bottlenecks'' are realized 
in the vicinity of T$_{MI}$ will
be the subject of future work.
We emphasize that whatever is
the actual mechanism responsible for the insulator-metal transition, the CO
phase appears to play a less important role in it. 

X-ray irradiation at low temperatures
destroys the CO state and converts the material to a
state which was previously found to be metallic in similar samples.
It appears that the mechanism for this low-temperature transition 
does not require the presence of the
phase distinct from both the FM and the
CO phases. On the other hand, it has been shown that 
x-rays do not strongly affect the CO state
above T$_{MI}$ while still converting the material to the
metallic state \cite{Casa}. Thus, it is possible that 
the insulator metal transition at the relatively
high temperature T$_{MI}$ and the low-temperature
photoinduced transition have different microscopic mechanisms. 

We also showed that the insulator-metal transition takes place in
samples with an extremely small low-temperature fraction of the FM phase.
The conducting phase in these samples must be of a highly filamentary 
character. We propose that lattice defects and associated strains play 
an important role in these samples.

We are grateful to M. E. Gershenson,
A. J. Millis, and D. Gibbs for important discussions. 
This work was supported 
by the NSF under Grant No. DMR-9802513, and by Rutgers University.
Work at Brookhaven National Laboratory was carried out under contract 
No. DE-AC02-98CH10886, US Department of Energy.


\begin{references}
\bibitem{saturation} Note that in Fig. 3(a) the peak intensity 
is shown, not the integrated intensity. 
The data of Fig 3(b) indicate that the integrated intensity plot
would give a more conventional ``order parameter'' curve with a rapid increase
in the vicinity of T$_{CO}$ and a slower variation at low temperatures.
\bibitem{difference} The small difference between the CO transition 
temperatures determined from the diffraction and the specific heat 
measurements is probably due to the temperature broadening of the 
CO transition. Note that there is no true divergence of the orbital correlation
length at T$_{CO}$ and therefore this temperature is not well defined. 
Another possible reason for the observed discrepancy is a small 
miscalibration in the temperature measurement.
\end{references}
\end{document}